\documentclass[11pt]{article}
\usepackage[margin=1in]{geometry}
\usepackage{graphicx}
\usepackage{soul}
\usepackage{courier}
\usepackage{slashed}

\usepackage{hyperref}
\hypersetup{
  pdfinfo={
  pdfproducer={},
  Title={Recent Results and Future Plans of the MoEDAL Experiment},
  Subject={},
  Author={Michael Staelens},
  }
}

\def\beq{\begin{equation}}
\def\eeq#1{\label{#1}\end{equation}}
\def\eeqn{\end{equation}}

\def\beqa{\begin{eqnarray}}
\def\eeqa#1{\label{#1}\end{eqnarray}}
\def\eeqan{\end{eqnarray}}

\let\bar=\overbar

\def\Dslash{\not{\hbox{\kern-4pt $D$}}}
\def\dslash{\not{\hbox{\kern-2pt $\del$}}}

\def\msb{{\bar{\ssstyle M \kern -1pt S}}}

\def\Title#1{\begin{center} {\Large {\bf #1} } \end{center}}
\def\Author#1{\begin{center} {\normalsize {\sc #1} } \end{center}}
\def\Institution#1{\begin{center} {\normalsize {\it #1} } \end{center}}
\def\Abstract#1{\noindent {\normalsize {\bf Abstract:} {\normalfont #1}}}
\def\Conference{\vspace{4mm}\begin{raggedright} {\normalsize {\it Talk presented at the 2019 Meeting of the Division of Particles and Fields of the American Physical Society (DPF2019), July 29--August 2, 2019, Northeastern University, Boston, C1907293.} } \end{raggedright}\vspace{4mm}}

\begin{document}

\Title{Recent Results and Future Plans of the MoEDAL Experiment}

\Author{Michael Staelens}
\Author{On behalf of the MoEDAL Collaboration}

\Institution{Department of Physics\\ University of Alberta, T6G 2E1 Edmonton, CANADA}

\Abstract{The Monopole and Exotics Detector at the LHC (MoEDAL) is a pioneering LHC experiment designed to search for anomalously ionizing messengers of new physics such as magnetic monopoles or massive (pseudo-)stable charged particles.  These are predicted to exist in a plethora of models beyond the Standard Model.  Deployed at Interaction Point 8 (IP8) along the LHC ring, MoEDAL has taken data at centre-of-mass energies of 8 and 13 TeV.  Its ground breaking physics program defines over 40 scenarios that yield potentially revolutionary insights into such foundational questions as: are there extra dimensions or new symmetries; what is the mechanism for the generation of mass; does magnetic charge exist; and what is the nature of dark matter?  MoEDAL's purpose is to meet such far-reaching challenges at the frontier of the field.  We present a summary of the MoEDAL detector and its latest results on magnetic monopole production at the LHC.  Afterwards, progress on the physics program \& installation of MoEDAL's Apparatus for the detection of Penetrating Particles (MAPP) subdetector will be discussed.}

\Conference

%
%

\section{Introduction}
The MoEDAL experiment [1], which now involves more than 70 scientists worldwide, was unanimously approved by CERN's research board to begin taking data in 2015.  Situated along the LHC ring at IP8, the MoEDAL detector searches for massive (pseudo-)stable, slow-moving, singly or multiply (magnetically) charged particles.  There are many theories Beyond the Standard Model (BSM) which contain particles of this nature, many of which are described in [2].  To date, MoEDAL has taken data in 8 and 13 TeV p-p collisions, as well as in pb-pb collisions.  Currently, MoEDAL places the world's best direct limits on multiply charged and spin-1 (vector-like) magnetic monopoles [3].  We expect MoEDAL's lead in this arena to continue during LHC's RUN-3.  In this paper, we present MoEDAL's latest results as well as our predicted limits for RUN-3.  At this time, MoEDAL also plans to deploy the first version of MAPP, MAPP-1, to take 30 fb$^{-1}$ of data available at IP8.  The plan for this new subdetector, accompanied by results from two benchmarking processes, is also outlined in this paper.

The upgraded MoEDAL detector will expand MoEDAL's physics program by including searches for new long-lived neutral particles (LLPs) as well as minimally ionizing particles such as minicharged particles (mCPs), particles with $Q<<e$.  Both types of particles emerge in many BSM theories as well.  One well-motivated example of minicharged particles is in dark sector models with a new $U'(1)$ gauge group [4].  LLPs arise in many well-studied supersymmetric scenarios, for example: gauge-mediated supersymmetry (SUSY)  [5,6], R-parity violating SUSY [6-9], stealth SUSY [10], and mini-split SUSY [11].  Additionally, they are predicted in neutral naturalness [12] and certain relaxion models [13].  LLPs are present in models of dark matter [14], baryogenesis [15], neutrino masses [16], and hidden valleys [17], predicting a wide range of LLP production and decay morphologies.  The 2012 discovery of the Higgs boson also opens the interesting opportunity for Higgs mixing portals that generally admit exotic decays into LLPs.  Both forms of exotica are phenomenologically rich, and present likely candidates for the next particle discovery. 

The MAPP detector will be placed in the UGCI gallery which is adjacent to the MoEDAL region at IP8.  UGCI is a generously sized cavern which allows for decay zones of up to 10 m, facilitating the search for new LLPs.  This region also allows for a range of detector positions ranging from $\sim$55 m to $\sim$25 m from IP8, corresponding to angles of $\sim$5$^{\circ}$ and $\sim$25$^{\circ}$ respectively.  Between the UGCI gallery and IP8 is at least 25m of rock (more than 60 nuclear interaction lengths of material), which provides a natural shielding from SM collision products.  There is also a 100 m overburden of rock protecting the MAPP detector from cosmic ray particles.  Currently, the plan is to deploy MAPP-1 at a distance of $\sim$55 m from IP8, 5$^{\circ}$ to the beam line, as shown in Figure 1.  This allows MAPP to exploit the forward biasing of several of the physics channels investigated thus far, gaining a better effective acceptance as a result.  

The aim of this paper is to briefly outline MoEDAL's expectations for magnetic monopole, mCP, and LLP searches during the LHC's RUN-3 with the upgraded detector.  I begin with an introduction to the MoEDAL detector and display MoEDAL's latest results on monopole production at the LHC.  These results also incorporated photon-photon fusion into the analysis, for the first time at the LHC.  The MAPP detector, which is broken down into two subdetectors, MAPP-mCP and MAPP-LLP, is described in Section 3.  Section 4 then presents two Hidden Sector models of interest to MAPP, namely, mCPs and new LL light scalars produced by meson decays.  Progress on simulating these models and the MAPP detector is given, including the expected sensitivity of MAPP-1 to these mCPs produced via the Drell-Yan mechanism during RUN-3.  Lastly, Section 5 concludes with some final remarks and future plans for the development of the MAPP detector.

\begin{figure}[htb]
\centering
\includegraphics[height=3in]{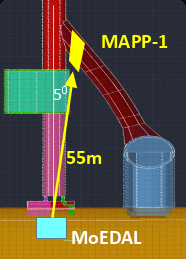}
\caption{\label{fig:1} Planned positioning of the MAPP-1 detector in the UGCI gallery, relative to the MoEDAL region.}
\end{figure}



\section{The MoEDAL Detector Today}

The MoEDAL detector is uniquely optimized to search for highly-ionizing particles, such as magnetic monopoles, dyons, highly electrically charged objects (HECOs), etc.  To achieve this, 3 main subdetectors are utilized: A large Nuclear Track Detector (NTD) Array, a magnetic trapping detector, and an array of TimePix chips.  A quick outline of each subdetector will be presented along with a description of their underlying analyses; refer to [18] for a full description of the detector design.  As mentioned previously, MoEDAL is placed at IP8 near the LHCb VELO region, covering a large geometric acceptance of 70$\%$\footnote{This is for the NTDs to detect at least one Drell-Yan pair produced monopole.}.  A \texttt{Geant4} model of the present MoEDAL detector is shown in Fig. 2.

\begin{center}
\begin{figure}
\centerline{\includegraphics[width=10cm, height=6cm]{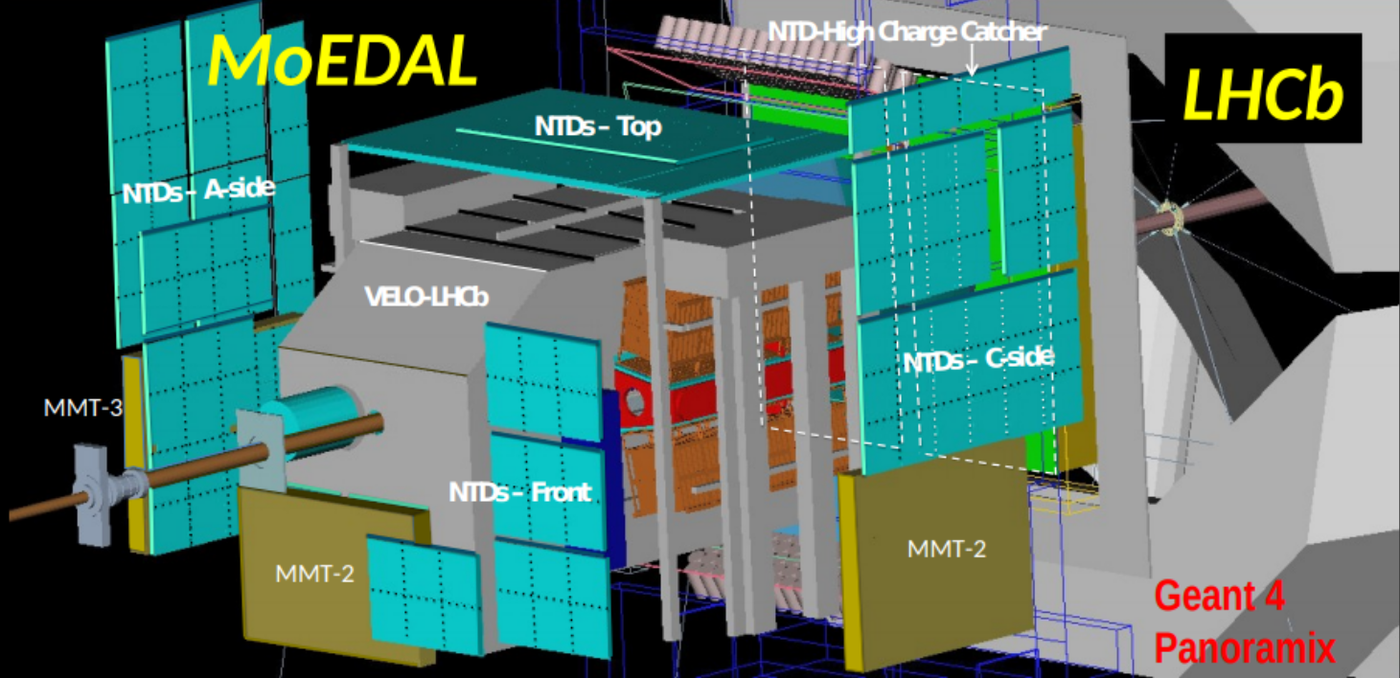}}
  \caption{\texttt{Geant4} Panoramix modeling of the present MoEDAL detector.}
\end{figure}
\end{center}

The MoEDAL detector operates passively, with the exception being the TimePix chip array which actively monitors backgrounds.  The purpose of the NTD system is to track the ionization paths of highly ionizing particles particles and measure their effective charge.  The minimum threshold for the NTDs  to register a particle's passage is $\frac{Z}{\beta}>5$, where $Z$ is the charge and $\beta$ is the velocity of the particle as a fraction of the speed of light. Thus, the NTDs are insensitive to Standard Model particles.  Etching the NTD sheets in a caustic solution, like NaOH, reveals the paths of highly ionizing particles.  Using the heavy-ion beams at NA61 as well as the NASA Space Radiation Facility (NSRL), the NTD material stacks are calibrated, making MoEDAL uniquely optimized to searching for highly ionizing signatures of new physics.  The etched material is then scanned by human assisted computerized optical scanning microscopes searching specifically for the signature tracks caused by highly ionizing particles, the size of which, is proportional to the charge of the particle.  In the future we intend to automate the process of scanning using machine learning techniques.  

Using the 794 kg of aluminum samples that form the magnetic trapping detector (MMT), MoEDAL can also directly detect magnetic charge.  The MMT volumes aim to slow down and trap magnetically charged particles that ionize as they pass through it.  Aluminum nuclei have a large, positive anomalous magnetic moment which facilitates monopole trapping.  With the NTD system alone, MoEDAL can place limits on various highly ionizing particle production channels.  However, with the combination of these two subdetectors, the reach of the MoEDAL detector is greatly enhanced.  The exposed NTD and MMT stacks are both removed yearly for analysis.  The MMT trapping volumes are sent to ETH Zurich where each sample is passed through a Superconducting QUantum Interference Device (SQuID) magnetometer.  The signal for a trapped magnetically charged particle would be a sharp rise in the current through the superconducting loop which remains constant after the MMT stack has passed through the SQuId.

Using the MMT subdetector which was exposed to 4.0 fb$^{-1}$ of 13 TeV proton-proton collisions at IP8, MoEDAL has placed some of the world's best limits on magnetic monopole production at the LHC via both Drell-Yan and photon-photon fusion processes [3].  In the analyses performed, MoEDAL considered monopoles with spins 0, 1/2, and 1, as well as assuming both $\beta$-dependent and -independent couplings.  Feynman-like diagrams for two of these channels are shown in Fig. 3.  From this search, MoEDAL excludes magnetic charges of $g>g_{D}=68.5 e$ in all samples and as a result, provides the best current limits on laboratory produced magnetic monopoles with $2g_{D}<g<5g_{D}$ and masses up to 2 TeV, as shown in Fig. 4.  The best limits for a magnetic charge of 1 $g_{D}$ are now held by ATLAS.  MoEDAL's expected contours for RUN-3 are also shown in Fig. 4.  For the full paper on MoEDAL's latest results see [3].

\begin{figure}[htb]
\centering
\includegraphics[height=1.5in]{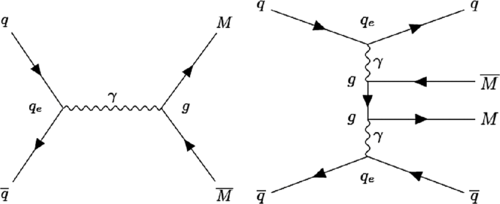}
\caption{\label{fig:3} Leading order Feynman-like diagrams for direct production of monopole pairs via the Drell-Yan (left) and $\gamma\gamma$-fusion (right) processes at the LHC.  A four-vertex diagram which is not shown, is also included for scalar and vector monopoles.}
\end{figure}

\begin{figure}[htb]
\centering
\includegraphics[height=3in]{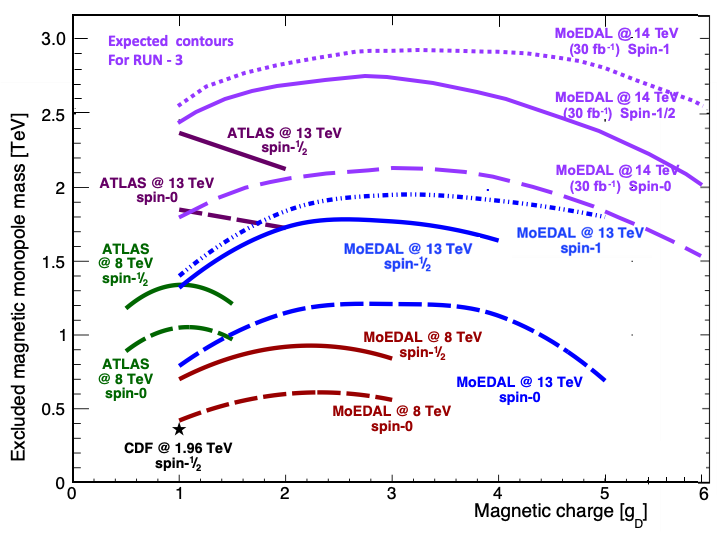}
\caption{\label{fig:4} Excluded magnetic monopole masses for pair produced monopoles at collider experiments, with MoEDAL’s latest results shown in blue and MoEDAL's expected RUN-3 results shown in purple (using $\beta$-independent couplings).}
\end{figure}


%
%

\section{MoEDAL's MAPP Subdetector}
The MAPP detector consists of 2 main subdetectors: A $\sim$150 m$^{3}$ LLP detector (MAPP-LLP) and a central 3 m$^{3}$ scintillation detector for mCPs (MAPP-mCP).  The compact central section of MAPP that forms MAPP-mcP is made of two co-linear sections, each with a cross-sectional area of 1 m$^{2}$ and comprised of 100 (10 x 10 cm) plastic scintillator sections.  Both sections are 1.5 m long and are further sub-divided into two 75 cm long bars.  Thus, a through-going particle from the IP will encounter a total of length 3 m (4 x 75 cm) through the scintillator bars.  The bars are each readout by a single PMT.  We will place all four PMTs in coincidence in order to essentially eliminate backgrounds from dark noise in the PMTs and radiogenic signals in the plastic scintillator or PMTs themselves.  The detectors are also protected from cosmic rays and from particle interactions in the surrounding rock by charged particle veto detectors. The MAPP-mCP detector must be able to produce a measurable signal which could be as small as a single photo-electron produced in the PMT, thus MAPP-1 will use large path length of scintillator developed by our IEAP group to have enhanced light output.  A protoype of MAPP-mCP was deployed at the UGCI gallery late in 2017 to acquire data and is currently being studied to assess the operating characteristics of MAPP-1.

The nested system of three detectors that surrounds the MAPP-mCP detector, forms MAPP-LLP.  Each of the three detector systems are formed of 5 faces butted against the forward veto plane, which acts as the sixth face and defines the start of the decay zone.  Each face acts as a type of hodoscope detector and is comprised of plastic scintillator strips which are arranged in an x-y configuration.  These are readout by silicon photomultipliers (SiPMs), allowing position and time-of-flight (ToF) tracking with expected resolutions of $\sim$500 ps and $\sim$1 cm, respectively.  The outer layer of the MAPP-LLP detector, which is roughly 10 x 5 x 3 m,  borders the fiducial volume of the MAPP-LLP system and defines the acceptance area.  The MAPP-LLP detector system also acts as a charged particle veto for the MAPP-mCP detector.  In addition, the MAPP-mCP detector is sandwiched between 3 1m$^2$ detector veto planes as further protection against charged particles that skim the edges of the scintillator bars comprising the  MAPP-mCP detector.  Lastly, the protected environment of the UGCI gallery also helps to provide a low radiation background situation for the MAPP detector.  A depiction of MAPP-1 deployed in the cavern is shown in Fig. 5, from two different perspectives. 



\begin{figure}[htb]
\centering
\includegraphics[height=3in]{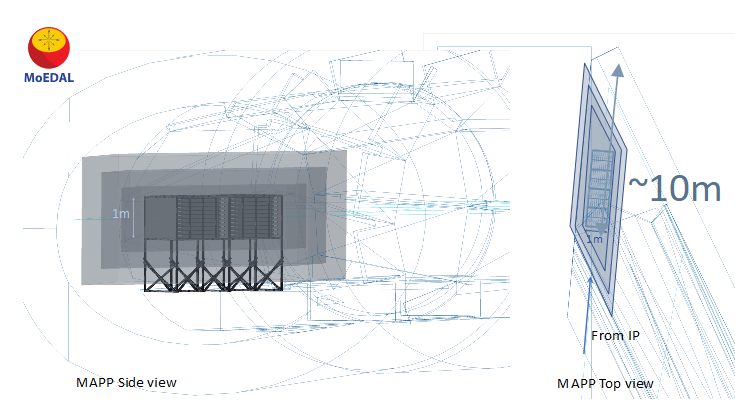}
\caption{\label{fig:5} A diagram of the MAPP-1 detector, depicting two different views.  A side view is given on the left, while a top-down view is shown on the right.  The top-down view highlights the angle offset between the two subdetectors.  MAPP-mCP is placed roughly normal to the 5$^{\circ}$ line, while MAPP-LLP is aligned with the cavern itself.}
\end{figure}

%
%

\section{MAPP-1 Physics Program Benchmarks}
The following two subsections each focus on a particular benchmark process, with the intention of illustrating the physics reach of the MAPP-1 detector.  Both models investigated involve a new dark sector that couples to the standard model through portal interactions.  The first process involves mCPs coupled to the SM via a vector portal (the Dark Photon), while the second involves a new long-lived scalar particle (the Dark Higgs) which couples to the SM Higgs.  As we will see, MAPP-1 expects sensitivity to both of these processes during the LHC's RUN-3.

%
%

\subsection{The Search for mCPs with MAPP-1}
Particles with an $Q<<e$, have been discussed in connection with the mechanism of electric charge quantization and possible non-conservation of electric charge [19].  As mentioned before, dark sector models with an additional U(1) gauge group typically contain new mCPs, a result of so-called kinetic mixing.  The possibility of detecting such particles with MAPP-1 is explored here, using an example scenario in which a new massless $U'(1)$ gauge field, the dark-photon ($A'_{\mu \nu}$), is coupled to the SM photon field, $B^{\mu \nu}$.  A new massive dark-fermion ($\psi$) of mass $M_{mCP}$, is also predicted, which is charged under the new U(1) field $A'$ with charge $e'$. The Lagrangian for the model is given by,

\begin{equation}
    \mathcal{L} = \mathcal{L}_{SM} - \frac{1}{4} A'_{\mu \nu}A'^{\mu \nu} + i\bar{\psi}(\slashed{\partial} +ie'\slashed{A'} + iM_{mCP})\psi -\frac{\kappa}{2}A'_{\mu \nu} B^{\mu \nu}.
\end{equation}
The last term contains the mixing, which one can eliminate by expressing the new gauge boson as, $A'_{\mu} \Rightarrow  A'_{\mu} + \kappa B_{\mu}$.  Applying this field redefinition reveals a coupling between the charged matter field $\psi$ to the SM hypercharge. The Lagrangian shown above then becomes:

\begin{equation}
    \mathcal{L} = \mathcal{L}_{SM} - \frac{1}{4} A'_{\mu \nu}A'^{\mu \nu} + i\bar{\psi}(\slashed{\partial} +ie'\slashed{A'} -i\kappa e' \slashed{B}  +iM_{mCP})\psi.
\end{equation}
It is now apparent that the field $\psi$ acts as a field charged under hypercharge with a minicharge (mCP) $\kappa e'$. This new minicharged matter field $\psi$ couples to the photon and Z$^0$ boson with a charge $\kappa e'\cos{\theta}_{W}$ and $– \kappa e' \sin{\theta}_{W}$, respectively. Expressing the fractional charge in terms of electric charge thus gives $\epsilon = \kappa e' cos{\theta}_{W}/e$.

\begin{center}
\begin{figure}
\centerline{\includegraphics[height=3in]{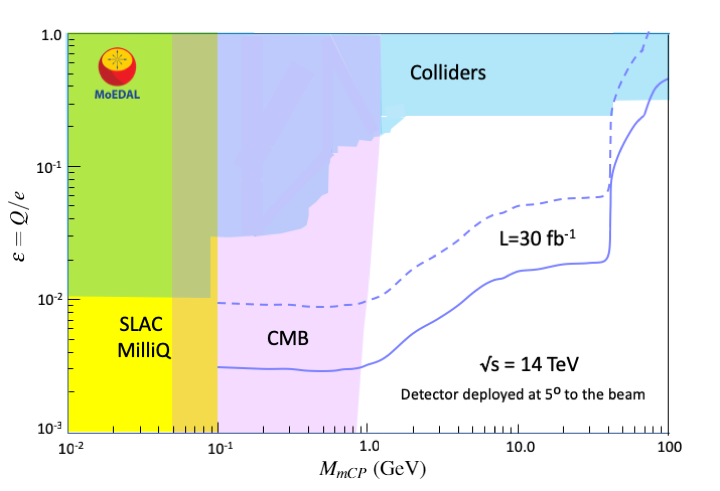}}
  \caption{The maximum reach of the MAPP-1 detector using 30 fb$^{-1}$ of data in LHC's RUN-3. The solid line represents that points at which 3 events would be observed assuming 100$\%$ efficiency and no background. The dotted line represents the case where 3 events are observed with an overall detector efficiency of 10$\%$.}
\end{figure}
\end{center}

Currently, the best direct limits on mCPs for 10$^{-1}$ to 10$^{-5}$ for $M_{mCP}$ $<$ 300 GeV have been derived from accelerator-based searches [20][21].  MilliQan [22], another future experiment dedicated to the search for mCP at the LHC, is also planned for LHC's Run-3.  The placement of this detector will be at 45$^\circ$ to the beam roughly 30m away from the CMS detector.  In addition, to collider searches, there are a number of indirect searches for mCPs using astrophysical systems  [23-25].  If one takes the astrophysical bounds at face value and combines them with direct searches, most of the parameter space for mCPs with masses 0.1 $\leq$ $M_{mCP}$ $\leq$ 100 GeV, is still unexplored, as shown in Figure 6.

In order to establish the potential of detecting minicharged particles at the LHC with MAPP-1's MAPP-mCP subdetector, we implemented the model presented here into \texttt{MadGraph5} [26] using \texttt{Feynrules} and \texttt{Mathematica}.  This model was validated by performing several comparisons with the literature available.  With this model we generated events at $\sqrt{s}=14$ TeV and assuming an integrated luminosity of 30 fb$^{-1}$, for various values of $M_{mCP}$ and $\epsilon$.  An in-house script analyzed these events and simulated MAPP-mCP, requiring a minimum of 3 observed mCP tracks through the detector.  This establishes MAPP-1's expected 95$\%$ confidence limit, which is also shown in Figure 6.  Here we assumed overall detector efficiencies of both 100$\%$ and 10$\%$.  MAPP-1 covers a significant portion of the free parameter space, which could be further improved by including $\psi$ couplings to the psion and $\Upsilon$ resonances.  With the factor of 10 increase in luminosity expected during HL-LHC, MAPP-1 can make further contributions to the search for minicharge.

%
%

\subsection{The Search for LLPs with MAPP-1}
To illustrate MAPP's physics reach for new LLPs, we used a simple benchmark scenario [27] where a dark Higgs mixing portal admits exotic inclusive $B \rightarrow X \varphi_{H}$ decays, where $\varphi_{H}$ is a light CP-even scalar that mixes with the SM Higgs, with a mixing angle of $\theta << 1$.  The particle lifetime depends on the degree of mixing.  We explore the decay $B \rightarrow K \varphi_{H}$ as a example to estimate MAPP-1's fiducial efficiency [28].  One possible, simple Lagrangian which includes this new dark Higgs mixing is given by the following [27], 

\begin{equation}
    \mathcal{L} = \mathcal{L}_{Kin} + \frac{\mu_{s}^{2}}{2} S^{2} - \frac{\lambda_{S}}{4!} S^{4} + \mu^{2} |H|^{2} - \lambda |H|^{4} - \frac{\epsilon}{2} S^{2} |H|^{2},
\end{equation}
where S is a real scalar field, H is a SM Higgs-like field, $\epsilon$ is the portal coupling, and $\lambda_{S}$ is a free parameter.  The final term contains the mixing between the SM Higgs and the new scalar, with the resulting physical fields being the SM Higgs and the dark Higgs.  Both fields acquire a non-zero VEV and the coupling between these two particles induces new Yukawa-like couplings between the dark Higgs and the SM fermions.  Thus, the signal we look for in MAPP-LLP is two charged lepton tracks originating from dark Higgs decays $\varphi_{H} \rightarrow l^{+}l^{-}$, in the fiducial volume of MAPP-1.  Each track must hit at least two consecutive planes to be detected in the MAPP-1 detector.  

Currently the best experimental limits on dark Higgs production at colliders come from CHARM [29] and LHCb [30,31].  In order to investigate MAPP-1's potential, we simulate the nominal detector geometry stated here and generate Monte-Carlo events at a collision energy of 14 TeV using \texttt{Pythia8} [32].  We explore a range of decay lengths ($c \tau$) and dark Higgs masses, and find that MAPP-1 has a maximum fiducial efficiency of $\sim 5 \times 10^{-4}$.  We are currently in the process of generating limit curves for this process, however we are still working out detailed simulations of the backgrounds and tracking efficiencies expected with MAPP-1 first.  We also note that, since the dominant production mechanism of $B \bar{B}$ production at the LHC is through gluon-gluon fusion in which the momenta of the incoming partons is strongly asymmetric in the laboratory frame.  The resulting center of mass energy of the produced $B \bar{B}$ pair is boosted along the direction of the higher momentum gluon and both B-hadrons are typically produced in the forward (or backward) direction.  Thus, MAPP-LLP also benefits from the 5$^{\circ}$ placement of MAPP-1 for the process shown here, and we expect this will be true for many mesonic decay modes to new light particles.  


%
%

\section{Conclusion}
MoEDAL's latest results and upgrade plans for the LHC's RUN-3 were presented, with particular attention paid to the new MAPP subdetector and two benchmarking processes considered.  The design and construction of MAPP-1 was also outlined.  Based on the current plan and results shown here, MAPP-1 is on schedule for installation by RUN-3 and we expect it will make a relevant contribution to the search for mCPs and LLPs at the LHC.  Presently, several other new physics channels available to MAPP-1 are also been investigated, while a few more a still being considered.  Additionally, we are completing a full \texttt{Geant4} simulation of the MAPP-1 detector and the material between it and IP8.  Lastly, detailed analyses of the backgrounds and LLP decay product tracks expected in MAPP-1 are also underway.  

\section*{Acknowledgements}
This work was supported by the grant NSERC-SAPPJ-2019-00040.

\end{document}